# RecBayes:
# Recurrent Bayesian *Ad Hoc* Teamwork in Large Partially Observable Domains

João G. Ribeiro[1,4], Yaniv Oren[2], Alberto Sardinha[3], Matthijs Spaan[2], Francisco S. Melo[1]
j.m.godinho.ribeiro@tue.nl

[1]**INESC-ID & Técnico Lisboa**
[2]**Delft University of Technology**
[3]**Pontifical Catholic University of Rio de Janeiro**
[4]**Eindhoven University of Technology**

## Abstract

This paper proposes RecBayes, a novel approach for *ad hoc* teamwork under partial observability — a setting where agents are deployed on-the-fly to environments where pre-existing teams operate — that never requires, at any stage, access to the states of the environment or the actions of its teammates. We show that by relying on a recurrent Bayesian classifier trained using past experiences, an *ad hoc* agent is effectively able to identify known teams and tasks being performed from observations alone. Unlike recent approaches such as PO-GPL (Gu et al., 2021) and FEAT (Rahman et al., 2023), that require at some stage fully observable states of the environment, actions of teammates, or both, or approaches such as ATPO (Ribeiro et al., 2023a) that require the environments to be small enough to be tabularly modelled (Ribeiro et al., 2023a), in their work up to $4.8K$ states and $1.7K$ observations, we show RecBayes is both able to handle arbitrarily large spaces while never relying on either states and teammates' actions. Our results in benchmark domains from the multi-agent systems literature, adapted for partial observability and scaled up to $1M$ states and $2^{125}$ observations, show that RecBayes is effective at identifying known teams and tasks being performed from partial observations alone, and as a result, is able to assist the teams in solving the tasks effectively.

## 1 Introduction

*Ad hoc* teamwork (Stone et al., 2010) studies how autonomous agents can be deployed to multi-agent environments where pre-established teams cooperate to perform tasks. When deployed on-the-fly, agents find themselves facing potentially unexpected situations, unable to pre-coordinate or pre-communicate with their new teammates. Under such conditions, agents are required to solve two main problems: (i) how to identify known teams and tasks being performed, to assist them accordingly and/or (ii) how adapt to unknown teammates and tasks without having the time to properly train from scratch. These two problems provide a distinct and challenging framing that distinguishes the setting from multi-agent reinforcement learning. For this reason, *ad hoc* teamwork has recently fallen under great attention by the multi-agent systems community (Mirsky et al., 2022), with some approaches addressing the first problem — how to identify and assist teammates which have been encountered before (Barrett et al., 2017; Gu et al., 2021; Rahman et al., 2023), and others addressing the second — how to assist teammates which have never been encountered before (Chen et al., 2020; Neves & Sardinha, 2022; Ribeiro et al., 2023b). This paper focuses on the first problem, where the agent encounters known teams and tasks, and does not consider the scenario where the agent must adapt to entirely new teammates or tasks.

Current approaches addressing the first problem often focus on leveraging an agent's past experiences to identify teams on-the-fly, relying on models to predict the behaviour of each team and computing a best-response policy accordingly. The problem with these methods is that they often rely on privileged information not always available, such as the full state of the environment or the real-time actions of its teammates (Barrett et al., 2017; Chen et al., 2020; Gu et al., 2021; Ribeiro et al., 2023b; Rahman et al., 2023). Two recent approaches, PO-GPL (Rahman et al., 2023) and FEAT (Fang et al., 2024), for instance, although exploring partially observable environments, still rely on the actions of teammates to identify the teams. Others, such as ODITS (Gu et al., 2021) deal with partial observable environments by explicitly pre-training the *ad hoc* agent in fully observable environments alongside each team, where both the states and the teammates' actions can be observed. Arguing that allowing an agent an explicit fully observable 'pre-training' stage dilutes the purpose of *ad hoc* teamwork under partial observability, other authors (Ribeiro et al., 2023a) show that by having human experts modelling past experiences under partial observability, it is possible to rely on them to identify teams and tasks on-the-fly, never requiring states or actions of teammates to be observable. However, besides requiring human experts with sufficient domain knowledge to create these models, a key problem with this approach is its tabular nature, which requires the environments to be small enough to be modelled, shown by the authors in their evaluation with environments up to $4.8K$ states and $1.7K$ observations.

To address these limitations — (i) the requirement of privileged information such as states of the environment or the actions of teammates observable at any stage and (ii) the inability to handle environments with arbitrarily large state and observation spaces, we propose a novel approach for *ad hoc* teamwork under partial observability named Recurrent Bayesian *Ad Hoc* Teamwork in Large Partially Observable Domains, or RecBayes for short. To the best of our knowledge, RecBayes is the first approach not only capable of identifying known teams and tasks being performed when deployed on-the-fly, but also do so without requiring, at any stage, the states of the environment or the actions of the teammates to be observable. We demonstrate the effectiveness of our approach in multi-agent environments containing up to $1M$ states and $2^{125}$ observations using two benchmark domains from the multi-agent systems literature. Our results show that our approach is not only effective at identifying known teams and tasks from observations alone and, consequently, efficient in assisting the teams solve the tasks. We believe the benefits of our approach introduce significant implications for future real-world applications, enabling *ad hoc* teamwork in more complex and dynamic environments.

## 2 Related Work

*Ad hoc* teamwork literature can be framed according to how it addresses three key sub-problems: (i) task identification, (ii) team identification and (iii) planning/execution (Melo & Sardinha, 2016). Seminal works (Stone & Kraus, 2010; Barrett & Stone, 2011) started by assuming that both the team and the task being performed were known beforehand, with planning and execution being their only focus, with team identification being later on approached by Chakraborty & Stone (2013). Following Melo & Sardinha (2016)'s formalisation of *ad hoc* teamwork, Barrett et al. (2017) later introduce the PLASTIC framework, which allows *ad hoc* agents to identify known teams by observing their actions in fully observable environments. The authors were then followed by Chen et al. (2020), which propose AATEAM, an approach employing an attention mechanism to weigh characteristics from known teams when facing unknown ones, and later on TEAMSTER, proposed by Ribeiro et al. (2023b) to efficiently adapt to new teams, continuously updating a shared world-model. To perform team identification, however, all three approaches require the *ad hoc* agent to (i) observe the actions of its teammates and (b) fully observe the states of the environment.

Barrett et al. (2017) where later followed by Ribeiro et al. (2021), Rahman et al. (2023) and Fang et al. (2024), who deal with these two limitations, however individually. Ribeiro et al. (2021) first propose BOPA, extending the PLASTIC framework for environments where the teammates' actions are not observable and the *ad hoc* agent is forced to rely only instead on the states to identify known teams, and Rahman et al. (2023) symmetrically propose PO-GPL, which extends the PLASTIC framework to environments where the *ad hoc* agent is unable to fully observe the states, but may observe the actions of its teammates. Similarly, Fang et al. (2024) propose FEAT, expanding PO-GPL to domains where sets of teams do not need to be created *a priori* but instead extracted from previous interactions by leveraging meta-learning techniques. As PO-GPL, FEAT also relies on the actions of its teammates to identify them on-the-fly.

Taking these limitations into account, Gu et al. (2021) later on propose ODITS, an approach that enables *ad hoc* agents to identify teams from partial observations if they are able to train with them in fully observable environments where both the states and their actions are observable. Arguing that fully observable 'pre-training' environments (i) may not always be available and (ii) greatly dilute the main goal of *ad hoc teamwork*, Ribeiro et al. (2023a) expand upon Gu et al. (2021) by proposing ATPO, which instead relies on human experts to create tabular models from previous interactions under partial observability, modelling not only the teams but also the tasks they might be performing. In a follow-up line of work (Ribeiro et al., 2024), the authors even show how these methods can be used in real-world partially observable domains where humans interact with robots. Even though these methods are able to identify teams on-the-fly from observations alone, they require the environments to have small enough state and observation spaces, given their tabular nature. This limitation is even showcased by the authors in their evaluation, with environments up to 4831 maximum states and 1731 maximum observations, and best-response policies were reported to take up to 12h to compute.

To the best of our knowledge, there has yet been an approach capable of identifying teams (and tasks) without requiring full observability of the environment, the actions of its teammates, as well as the capacity to deal with domains with arbitrarily large state and observation spaces. Taking all the relevant literature into account and these limitations, we expand upon Gu et al. (2021) and Ribeiro et al. (2023a), proposing an approach capable of identifying teams and tasks directly from partial observations in arbitrarily large domains. Our approach, named Recurrent Bayesian *Ad Hoc* Teamwork in Large Partially Observable Domains, or RecBayes for short, shows that by relying on trajectories collected during previous interactions, always under partial observability, a recurrent Bayesian classifier can be effectively trained to approximate a probability distribution over all teams and tasks. Afterwards, as commonly done in most *ad hoc* teamwork literature (Barrett et al., 2017; Chen et al., 2020; Gu et al., 2021; Ribeiro et al., 2023b; Rahman et al., 2023; Ribeiro et al., 2023a), this distribution can be used in combination with a set of best-response policies associated with the known teams and tasks, computed using traditional reinforcement learning approaches.

## 3 Preliminaries

We frame *ad hoc* teamwork according to Melo & Sardinha (2016) and Ribeiro et al. (2023a), dealing not only with team identification, but also task identification. We consider partially observable domains where *ad hoc* agents are placed in an $\epsilon$ with an underlying team performing an underlying task. The *ad hoc* agent is then given the goal of identifying the most-likely team-task combination $k$ from a set of $K$ possible team-tasks, act accordingly to efficiently assist the team in solving the task. Ribeiro et al. (2023a) achieve this by explicitly modelling each team-task $k \in \{1, \ldots, K\}$ and respective environment $\epsilon^k$ as a POMDP $m^k = (\mathcal{X}^k, \mathcal{A}, \mathcal{Z}, P^k, O^k, r^k)$, where $\mathcal{X}^k$ represents the state space, $\mathcal{A}$ the *ad hoc* agent's action space, required equal for all team-tasks, $\mathcal{Z}$ the *ad hoc* agent's observation space, required equal for all team-tasks, $P^k$ the transition probabilities modelling both the world's dynamics as well as the behaviour of the team, $O^k$ the observation probabilities, $r^k$ the reward function.

Having set up a set of team-tasks models $\mathcal{L} = \{m^1, \ldots, m^K\}$, containing all $K$ POMDPs, Ribeiro et al. (2023a) set up a respective best-response policy library $\Pi = \{\pi^1, \ldots, \pi^K\}$, a common practice in *ad hoc* teamwork, which the authors compute relying on the PERSEUS algorithm (Spaan & Vlassis, 2005).

Then, when deployed at evaluation time to an unknown environment $\epsilon$, without knowing its underlying team and task $k$, the *ad hoc* agent starts by defining a prior distribution $p_0(k)$ for each model, which it then updates in each time step using the observation $z_{t+1}$ and the *ad hoc* agent's own action $a_t$ as evidence, i.e.

$$p_{t+1}(k) = p_t(k) \cdot \frac{1}{\rho} \cdot \sum_{x,y \in \mathcal{X}^k} O^k(z_{t+1} \mid y, a_t) \cdot P^k(y \mid x, a_t) \cdot b_t^k(x) \cdot \pi_t(a_t \mid h_t) \quad (1)$$

where $\rho$ is a normalisation constant, $b_t^k(x)$ the belief that the agent is in state $x$ at time step $t$ if the POMDP is $m^k$, and $\pi_t(a_t \mid h_t)$ the probability of the agent choosing action $a_t$ given the history $h_t = \{a_0, z_1, r_1, \ldots, a_{t-1}, z_t, r_t\}$.

Finally, the authors use the probability of each model, $p_t(k)$ as a weight to compute the agent's final policy, i.e.

$$\pi_t(a \mid h_t) = \sum_{k=0}^{K} \pi^k(a \mid h_t) \cdot p_t(k) \quad (2)$$

## 4 Recurrent Bayesian AHT in Large Partially Observable Domains

We now describe our proposed recurrent Bayesian algorithm for *ad hoc* teamwork in large partially observable domains. As shown by Equation 1, identifying the most-likely team-task is possible via Bayesian updates if the agent's has access to the transition and observation probabilities of the environment. Since these tabular models are unfeasible to set up for environments with large state and observation spaces, our hypothesis is trajectories collected during previous interactions with individual team-tasks $k$ may contain enough information to approximate the Bayesian update from Eq. 1 by framing it as a recurrent classification problem.

---

**Algorithm 1** Trajectory Collection for Previous Team-Task Combinations

**Input:**
    Environments where each team-task is present $\{\epsilon^1, \ldots, \epsilon^K\}$
    Number of trajectories to collect per environment $T$
    Max trajectory length $L$
**Output:**
    Trajectory buffers $\{traj^1, traj^2, \ldots, traj^K\}$
    Trained policies $\{\pi_{\psi^1}, \ldots, \pi_{\psi^K}\}$
**for** $k = 1$ to $K$ **do**
    Initialize policy $\pi_{\psi^k}$ and trajectory buffer $traj^k$
    **for** $t = 1$ to $T$ **do**
        Reset $\epsilon^k$ and trajectory $\tau = \emptyset$
        **for** $l = 1$ to $L$ **do**
            Observe $\phi(z_t)$
            Execute action $a_l \sim \pi_{\psi^k}(\phi(z_t), h_t)$
            Observe $\phi(z_{l+1})$ and reward $r_{l+1}$
            Append $(a_l, z_{l+1}, r_{l+1})$ to $\tau$
        **end for**
        Append $\tau$ to $traj^k$
        Update policy $\pi_{\psi^k}$ using trajectories in $traj^k$
    **end for**
**end for**
**return** $\{traj^1, \ldots, traj^K\}, \{\pi_{\psi^1}, \ldots, \pi_{\psi^K}\}$

---

**Algorithm 2** Recurrent Bayesian Classifier Training

**Input:**
    Trajectory buffers $\{traj^1, traj^2, \ldots, traj^K\}$, respectively labelled $k \in \{1, 2, \ldots, K\}$
    Number of epochs $E$
    Learning rate $\alpha$
    Batch size $B$

**Output:**
    Trained Recurrent Bayesian Classifier $\hat{p}_\theta$ with parameters $\theta$

Initialize classifier $\hat{p}_\theta$ with random parameters $\theta$
Create labelled dataset $\mathcal{D} = \{(\tau_1, k_1), \ldots, (\tau_N, k_N)\}$ using $\{traj^1, traj^2, \ldots, traj^K\}$
**for** epochs = 1 to $E$ **do**
    Sample batch $\{(\tau_1, k_1), \ldots, (\tau_B, k_B)\}$ from $\mathcal{D}$
    Compute logits $\{z_1, \ldots, z_B\} = \hat{p}_\theta(\{\tau_1, \ldots, \tau_B\})$
    Compute probabilities $\{p_1, \ldots, p_B\} = \text{softmax}(\{z_1, \ldots, z_B\})$
    Compute categorical cross-entropy loss $\mathcal{L}(\theta) = -\frac{1}{B} \sum_{b=1}^{B} \sum_{k=1}^{K} p_b^k \log(p_b^k)$
    Compute gradient $\nabla_\theta \mathcal{L}(\theta)$
    Update parameters $\theta \leftarrow \theta - \alpha \nabla_\theta \mathcal{L}(\theta)$
**end for**
**return** Trained Recurrent Bayesian Classifier $\hat{p}_\theta$

Specifically, when interacting with previous team-tasks $k \in \{1, \ldots, K\}$, we collect $T$ trajectories $traj^k = \{\tau_1^k, \ldots, \tau_T^k\}$, one for each team-task, with $\tau_i^k = (a_0, z_1, a_1, \ldots, a_{L-1}, z_L)$, $a_{t-1}$ denoting the actions executed by the *ad hoc* agent and $z_t$ its partial observations. After collecting the trajectories, we train parametrised recurrent Bayesian classifier $\hat{p}_\theta$ using the labelled dataset $\mathcal{D} = \{(\tau_1, k_1), \ldots, (\tau_N, k_N)\}$, where each datapoint $(\tau_1, k_1)$ represents a trajectory and its associated team-task. After training, $\hat{p}_\theta$ becomes optimised to approximate a distribution over all teams and tasks $k \in \{1, \ldots, K\}$, when recurrently given evidences $(a_{t-1}, z_t)$ as input. Algorithm 1 details the process for collecting the trajectories when interacting with different teams performing potentially different tasks, alongside the training of the respective best-response policies for each one. Algorithm 2, then details how these trajectories can be used to train the recurrent Bayesian classifier $\hat{p}_\theta$ by formulating as a recurrent classification problem. A natural consequence of this formulation is that we can relax the requirement made by Ribeiro et al. (2023a) that the observation's explicit tabular entry $z_t$ is required, being instead described by an arbitrary set of features $\phi(z_t)$.

**Algorithm 3** Recurrent Bayesian *Ad Hoc* Teamwork in Large Partially Observable Domains

**Input:**
    Unknown environment $\epsilon$
    Trained Bayesian classifier $\hat{p}_\theta$
    Set of known policies $\{\pi_{\psi^1}, \ldots, \pi_{\psi^K}\}$ for each team-task
    Number of evaluation episodes $E$

Initialize prior distribution $p_0$
**for** *episode* = 1 to $E$ **do**
    Reset environment $\epsilon^k$
    **while** not terminal **do**
        Observe $\phi(z_t)$
        Compute policy $\pi_t(a \mid \phi(z_t)) = \sum_{k=0}^{K} \pi_{\psi^k}(a \mid \phi(z_t), h_t^k) \cdot p_t$
        Execute action $a_t \sim \pi_t$
        Observe $\phi(z_{t+1})$ and reward $r_{t+1}$
        Update priors $p_{t+1} = \hat{p}_\theta(a_t, z_{t+1}; h_t)$
    **end while**
**end for**

Given the set of trained best-response policies $\{\pi_{\psi^1}, \ldots, \pi_{\psi^K}\}$, a common method in the current *ad hoc* teamwork literature (Barrett et al., 2017; Chen et al., 2020; Gu et al., 2021; Ribeiro et al., 2023b; Rahman et al., 2023; Ribeiro et al., 2023a), alongside the trained recurrent Bayesian classifier $\hat{p}_\theta$ able to compute a distribution over all known team-tasks, the *ad hoc* agent's final policy can be computed by adapting Equation 3 as it follows:

$$\pi_t(a \mid \phi(z_t)) = \sum_{k=0}^{K} \pi_{\psi^k}(a \mid \phi(z_t), h_t^k) \cdot \hat{p}_\theta(k \mid \phi(z_t), h_t) \tag{3}$$

where $h_t^k$ represents the a latent recurrent state kept by policy $\pi_{\psi^k}$ and $h_t^k$ the latent recurrent state kept by the Bayesian classifier $\hat{p}_\theta$. Putting it all together, Algorithm 3 summarises the algorithm for performing *ad hoc* teamwork alongside its action selection at evaluation time.

## 5 Experimental Setup

We now describe both evaluation procedure. In our evaluation, we aim to answer two main main research questions:

- **RQ1**: Is RecBayes efficient in assisting the teams solving the tasks?
- **RQ2**: Is RecBayes effective in identifying the teams and tasks from observations alone?

We follow the evaluation procedure for *ad hoc* teamwork proposed by Stone et al. (2010), which first evaluates a team of original teammates and afterwards replaces one of the members, at random, with the *ad hoc* agent to be evaluated. A single trial consists on running an episode where an *ad hoc* agent is deployed on-the-fly and stays with the team until the task is solved. When the episode is over, we register the total number of steps it took to complete. Each experiment, defined by one agent, deployed into one team performing one task, in one domain, for one world-size consists on 16 trials.

We now describe our domains, different teams, different tasks and baselines.

### 5.1 Domain Descriptions

We evaluate our approach in two benchmark domains from the *ad hoc* teamwork literature — the Level-Based Foraging (LBF) domain and the Predator Prey (PP) domain (figure 1), scaled up to provide a large enough state space and adapted for partial observability, as although these domains are commonly used benchmarks within both the multi-agent systems and *ad hoc* teamwork literature, they are often instantiated in relatively simple or small variations, such as fully observable $5 \times 5$ worlds. To showcase the effectiveness of our approach in arbitrarily large state spaces, we mitigate this by conducting experiments in variations of the environments ranging from the $7 \times 7$ as the smallest ones, containing 117649 total states, up to the $10 \times 10$, contain a total of 970200 states as the largest ones. Similarly, we introduce partial observability by adapting the observation function from Santos et al. (2022) which fixes a limited field-of-view around the agent, containing several channels for identifying other agents, walls and other information such as the levels. For simplicity, in all experiments and world sizes, we keep a field-of-view of size $5 \times 5$, providing a large and non unidimensional observation space containing up to $2^{125}$ different observations. Figure 2 showcases two examples in an environment from the Level-Based Foraging domain.

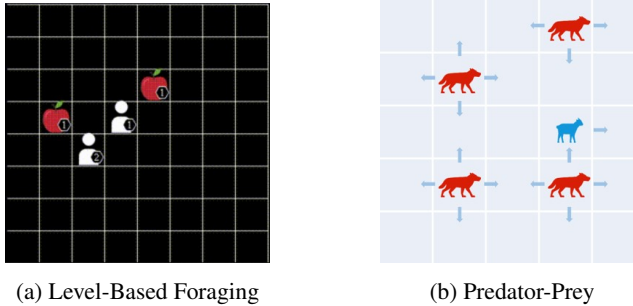

(a) Level-Based Foraging  (b) Predator-Prey

Figure 1: Visualisation of two example states in each of our benchmark environments, (a) Level-Based Foraging and (b) Predator-Prey.

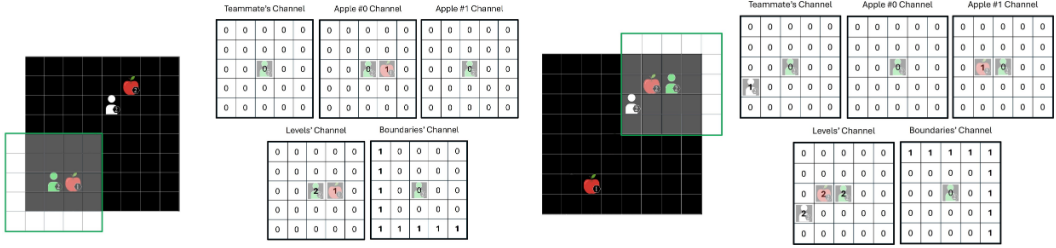

Figure 2: Example tuple observations of shape $5 \times 5 \times 5$. Five different channels of size $5 \times 5$ provide the agent a limited field-of-view which detects both entities and other information in the near environment. Each other agent is given a separate channel since indexes are relevant in these domain, used for instance for stalemate resolution or other decision processes.

### 5.2 Identification Sub-Problems

In our evaluation, we conduct in both domains three distinct sets of experiments — (i) **Team Identification**, where the task is fixed and the *ad hoc* agent has to identify and assist the correct team, (ii) **Task Identification**, where the team is fixed as the *ad hoc* agent has to identify and assist in performing the correct task and (iii) **Task & Teammate Identification**, where neither the task nor the team is fixed, requiring the *ad hoc* agent identify and assist the team in completing the tasks. We now describe the possible teams and tasks.

**Teams** Teams from both domains can have three different strategies. The first strategy (named Greedy) finds the closest target and moves towards it without taking into account any obstacles in the field. This strategy does not make any assumptions regarding its teammates. The second strategy (named Teammate Aware), picks the target with the highest value, i.e. highest level in the case of LBF and prey closer to all predators in the case of PP, and moves towards it taking into account obstacles such as its teammates. This strategy assumes its teammates share the same decision-making process. Finally, the third strategy (named Probabilistic Destinations), also selects the highest value target taking into account obstacles in the field to avoid collisions, but searches for the positions of its teammates to computes a plan which encircles the target to avoid collisions and, in the case of the Predator-Prey domain, the prey escaping. As with the second team strategy, this one also assumes the teammates share the same decision-making.

**Tasks** In both domains, we define tasks by the cardinal directions assigned to each individual team member when approaching a target. For example in the Level-Based Foraging domain, where agents must coordinate to pick-up apples spread out throughout the field, in each specific task, only a single agent may attempt to forage the apple from the north, another from the south, another from the west and another from the east. We also allow an additional task where there are no assigned coordinates.

### 5.3 Baselines

We evaluate a total of six baselines in the role of the *ad hoc* agent, two ablations of our approach, (i) one relying on a set of model-free reinforcement learning policies (**RecBayes-MF**) and (ii) another relying on a set of model-based reinforcement learning policies (**RecBayes-MB**), (iii) a teammate following the same policy as the original team (**Original Teammate**), (iv) an oracle model-free RL agent which knows the correct team-task beforehand (**Model-Free RL**), (v) an oracle model-based RL agent which knows the correct team-task beforehand (**Model-Based RL**) and (vi) a **random policy**.

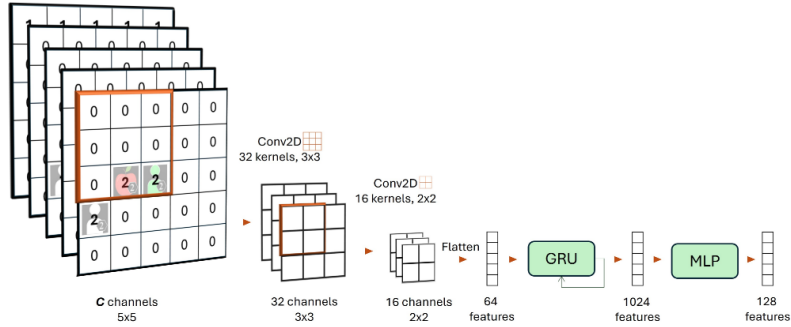

Figure 3: Encoder (or input) networks used both by the Bayesian classifier $\hat{p}_\theta$ and policies $\pi_\psi$ for automatic feature extraction. Observations $\phi(z_t) \in \mathbb{R}^{(5\times5\times5)}$ undergo 2D Convolutional layers, extracting a latent observation $\hat{z}_t \in \mathbb{R}^{64}$, followed by Gated Recurrent Unit and an additional feedforward network (or MLP) with two hidden layers, extracting a final latent state $\hat{x}_t \in \mathbb{R}^{128}$.

We evaluate both the model-free and model-based approaches to demonstrate that our approach is viable both with model-free and model-based RL, which alternate in their domination as the state-of-the-art in different domains. Even though our approach is not coupled with any particular algorithm, we chose for this evaluation PPO (Schulman et al., 2017) as our model-free algorithm and DreamerV3 (Hafner et al., 2023) as our model-based one, two popular RL algorithms, with their hyperparameters fine-tuned using Optuna (Akiba et al., 2019) and fully available in file `hyperparameters.yaml` in our repository at \*\*Anonymised for Review\*\*.

### 5.4 Model Implementations

Both our Bayesian Classifier $\hat{p}_\theta$ and policies $\pi_\psi$ require an input module capable of dealing with partial tuple-based observation space shape $5 \times 5 \times 5$.

To extract strong latent representations than can be used by both the classifier and best-response policies, a natural design choice is to implement these input modules as recurrent convolutional encoders, which we follow through in this work. Figure 3 provides an overview of the architecture, manually fine-tuned. Alongside these design choices, all hyperparameters are fully available in file `hyperparameters.yaml` in our repository.

## 6 Results

We now present the results from our experiments. We conduct our experiments with all six agents, on all three sets, on both domains, on both space sizes ($|\mathcal{X}| = 117649, |\mathcal{Z}| = 2^{125}$ and $|\mathcal{X}| = 970200, |\mathcal{Z}| = 2^{125}$). Agents relying on deep learning models are evaluated three total times with different parameter initialisations. In total, for all variations and baselines, 12544 trials were conducted in our experiments.

We now break down our results by research question.

### 6.1 Research Question #1: Is RecBayes efficient in assisting the teams solving the tasks

To answer our first research question, we compare the results obtained by RecBayes against the other baselines. Table 1 presents the average numbers of steps to complete an episode for all baselines on all settings. As done by Ribeiro et al. (2023a), to better analyse the impact of each agent, we normalise the results between the ones achieved Original Teammate and the Random Policy. This way, if an agent has a relative performance above $1.0$, it means it has outperformed the Original Teammate, while relative performances below $0.0$ mean the agent has underperformed the Random Policy. We consider a relative performance above $0.90$ to be near optimal. Table 2 presents the normalised results.

Table 1: Average number of steps to solve an episode in all experiments, the smaller the better. For each baseline, team, task, domain, world-size, we repeat trials 16 times. Agents relying on deep learning models (Model-Free, Model-Based, RecBayes-MF and RecBayes-MB) are evaluated two additional times to account for three total parameter initialisations. In total, for all variations and baselines, $12544$ trials were conducted in our experiments.

|  | Original Teammate | | Model-Free RL (Knows Task/Team) | | Model-Based RL (Knows Task/Team) | | RecBayes MF | | RecBayes MB | | Random Policy | |
|---|---|---|---|---|---|---|---|---|---|---|---|---|
| **Level-Based Foraging (7x7)** | | | | | | | | | | | | |
| Task Identification | 6.87 | (±1.86) | 7.56 | (±2.42) | 8.13 | (±5.49) | 13.02 | (±10.47) | 15.35 | (±17.15) | 222.6 | (±222.73) |
| Team Identification | 8.57 | (±8.77) | 7.68 | (±5.06) | 7.31 | (±4.91) | 8.85 | (±7.5) | 10.24 | (±8.38) | 86.37 | (±82.44) |
| Task & Team Identification | 7.69 | (±6.3) | 7.62 | (±3.92) | 7.73 | (±5.24) | 11.96 | (±9.12) | 14.06 | (±11.43) | 168.11 | (±192.19) |
| **Predator-Prey (7x7)** | | | | | | | | | | | | |
| Task Identification | 9.55 | (±4.05) | 11.22 | (±6.1) | 9.79 | (±3.81) | 21.36 | (±15.14) | 17.34 | (±9.91) | 270.66 | (±280.81) |
| Team Identification | 16.69 | (±18.6) | 10.19 | (±7.16) | 10.11 | (±6.55) | 10.71 | (±7.47) | 10.64 | (±6.83) | 120.48 | (±123.13) |
| Task & Team Identification | 12.61 | (±13.04) | 10.71 | (±6.67) | 9.95 | (±5.35) | 16.75 | (±11.62) | 15.67 | (±9.28) | 209.35 | (±241.46) |
| **Level-Based Foraging (10x10)** | | | | | | | | | | | | |
| Task Identification | 9.34 | (±2.98) | 33.67 | (±50.32) | 21.52 | (±25.42) | 66.53 | (±90.49) | 33.86 | (±38.88) | 411.84 | (±266.54) |
| Team Identification | 11.48 | (±7.37) | 15.65 | (±21.42) | 15.15 | (±10.68) | 19.27 | (±18.59) | 25.88 | (±19.86) | 230.22 | (±187.08) |
| Task & Team Identification | 10.27 | (±5.44) | 26.69 | (±42.5) | 18.79 | (±20.69) | 37.43 | (±40.61) | 35.47 | (±41.48) | 338.96 | (±253.98) |
| **Predator-Prey (10x10)** | | | | | | | | | | | | |
| Task Identification | 12.11 | (±4.91) | 17.52 | (±10.25) | 16.27 | (±8.2) | 78.65 | (±185.69) | 59.38 | (±85.73) | 457.65 | (±293.54) |
| Team Identification | 42.23 | (±73.61) | 31.75 | (±29.52) | 20.17 | (±15.83) | 32.71 | (±46.42) | 18.77 | (±12.42) | 271.11 | (±242.77) |
| Task & Team Identification | 25.02 | (±50.58) | 25.0 | (±23.64) | 17.94 | (±12.23) | 26.29 | (±19.69) | 34.41 | (±24.68) | 353.03 | (±281.9) |

The first observation we can make is that, as expected, the Original Teammates was never significantly outperformed. In fact, we can observe that it was matched by both RL baselines in all cases, which validates their proper set up for each specific team-task. We can also observe that there are no statistically significant differences between the Model-Free and Model-Based RL agents in all 12 cases.

Table 2: Results from Table 1 normalised between the ones obtained by the Original Teammate and the Random Policy, the higher the better. Values equal and above $1.0$ (highlighted) mean an agent has matched or outperformed the Original Teammate, while values below $0.0$ mean an agent performed worse than the random policy.

|  | Original Teammate | Model-Free RL (Knows Task/Team) | Model-Based RL (Knows Task/Team) | RecBayes MF | RecBayes MB | Random Policy |
|---|---|---|---|---|---|---|
| **Level-Based Foraging (7x7)** | | | | | | |
| Task Identification | 1.00 | 0.99 | 0.99 | 0.97 | 0.96 | 0.00 |
| Team Identification | 1.00 | **1.01** | **1.02** | 0.99 | 0.98 | 0.00 |
| Task & Team Identification | 1.00 | 1.00 | **1.00** | 0.97 | 0.96 | 0.00 |
| **Predator-Prey (7x7)** | | | | | | |
| Task Identification | 1.00 | 0.99 | **1.00** | 0.95 | 0.97 | 0.00 |
| Team Identification | 1.00 | **1.06** | **1.06** | **1.06** | **1.06** | 0.00 |
| Task & Team Identification | 1.00 | **1.02** | **1.01** | 0.98 | 0.99 | 0.00 |
| **Level-Based Foraging (10x10)** | | | | | | |
| Task Identification | 1.00 | 0.94 | 0.97 | 0.86 | 0.94 | 0.00 |
| Team Identification | 1.00 | 0.98 | 0.98 | 0.96 | 0.93 | 0.00 |
| Task & Team Identification | 1.00 | 0.95 | 0.97 | 0.92 | 0.92 | 0.00 |
| **Predator-Prey (10x10)** | | | | | | |
| Task Identification | 1.000 | 0.98 | 0.991 | 0.85 | 0.89 | 0.00 |
| Team Identification | 1.00 | **1.05** | **1.10** | **1.04** | **1.10** | 0.00 |
| Task & Team Identification | 1.00 | **1.00** | **1.02** | 0.99 | 0.97 | 0.00 |

We then analyse the performances of RecBayes's ablations. RecBayes-MF was able to achieve near-optimal performance in 10 out of the 12 cases and RecBayes-MB in 11 out of the 12 cases. All three non-optimal cases were related with Task Identification, suggesting one of two hypothesis — (a) identifying the correct task is harder than identifying the correct team or (b) not having full certainty on the correct task has a higher impact on performance than not having full certainty on the correct team — which we address in the next sub-section. To conclude, we can positively say from these results that RecBayes is efficient in assisting the teams in solving the tasks.

### 6.2 Research Question #2: Is RecBayes effective in identifying the correct team-task from observations alone?

To answer our second research question, we record the RecBayes's output probabilities $p_t(k)$ over the episode's time steps, computed by the Bayesian Classifier $\hat{p}$. Figure 4 showcases these probabilities over the course of the interactions, for all three sets, Task Identification, Team Identification and Task & Team Identification. We consider the identified team-task as the most-likely team-task, or $k = \arg\max_{k \in \{1,...,K\}} p_t(k)$.

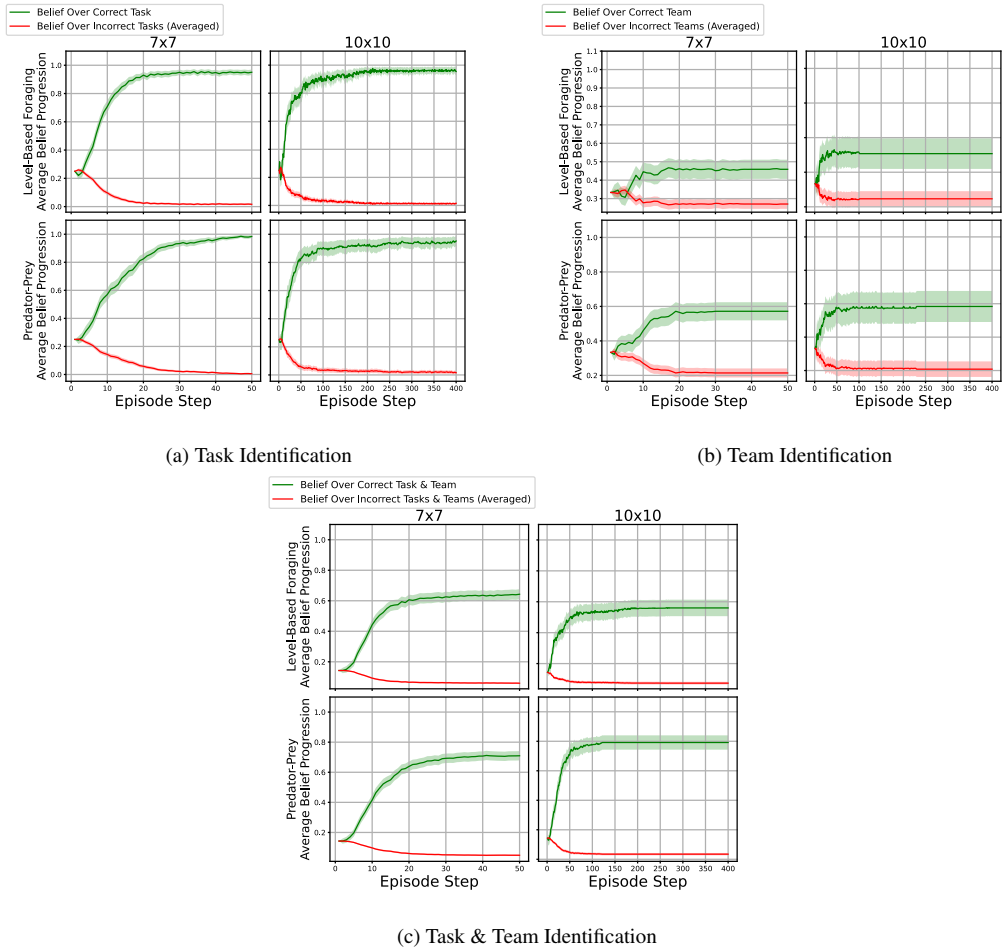

(a) Task Identification

(b) Team Identification

(c) Task & Team Identification

Figure 4: Identification of the correct team and task by RecBayes's Bayesian Classifier $\hat{p}$. In each timestep, $\hat{p}$ outputs a probability distribution over all team-tasks, with each entry representing the belief over its respective team-task. The belief for the correct team-task is shown in green, while the average sum of the incorrect ones is shown in red.

As we can see from these results, our proposed framing of team-task identification under partial observability as a sequence classification problem was effective in identifying the correct team-task on-the-fly. We also observe that tasks are more easily identified than teams. This answers our previous question regarding why RecBayes achieved a slightly lower performance on three Task Identification cases, while in Team Identification is was always near-optimal. In figure 4, however, we can clearly observe that identifying the correct task is easier than identifying the correct team, which seems to suggest that having higher certainty on the correct task is more important than having higher certainty on the correct team. Nonetheless, all team and tasks were in the end correctly classified, demonstrating our proposed Bayesian classifier effectiveness.

## 7 Conclusion

We propose RecBayes, a novel recurrent Bayesian approach for *ad hoc* teamwork in large partially observable domains which unlike prior approaches never requires at any stage access to the states of the environment nor actions of the teammates. We show that that by relying only on partially observable trajectories, collected by interacting with prior teams where neither the actions of the teammates nor the states of the environment are observable, a recurrent Bayesian classifier can be effectively trained to identify known teams performing tasks on-the-fly. Our evaluation, conducted in domains with environments with up to nearly $1M$ states and $2^{125}$ observations, where both teams and tasks are varied, has demonstrated that RecBayes is not only effective at identifying the correct team and task from a set of known teams and tasks, but also capable of assisting the teams in solving the tasks. We believe these results have significant implications regarding the development of *ad hoc* agents for real-world applications, consequently opening possible lines of future work.